\documentclass[preprint,review,12pt]{elsarticle}
\usepackage{amssymb}
\journal{Computational Materials Science}

\begin{document}

\begin{frontmatter}



\title{Ab initio  calculations for the tetragonal PbZr$_{0.5}$Ti$_{0.5}$O$_3$}


\author{Renata Bujakiewicz-Koronska}

\address{Institute of Physics, Pedagogical University, ul. Podchorazych 2, 
30-084 Krakow, Poland}

\begin{abstract}
\textit{Ab initio} studies of structural, elastic and electronic properties of the tetragonal perovskite-type  PbZr$_{0.5}$Ti$_{0.5}$O$_3$ are presented using the pseudo-potential plane wave method within the density functional theory in generalized – gradient approximation. The calculated equilibrium lattice parameters remain in a good agreement with the available experimental data. The bulk modulus obtained from the Birch-Murnaghan equation of state is calculated as B$_0$=170 GPa, and the gap energy E$_g$=2.1 eV-3.5 eV. The some differences between calculated and nominal charges exist for all atoms. The biggest ones are on the Pb ions.  They are caused by hybridization of the Pb 6\textit{s}  and O 2\textit{p} states. The influence of the strain on the averaged over directions Young modulus in the 0.1\%-0.3\%  range was studied.
\end{abstract}


\end{frontmatter}

% \linenumbers

\section{Introduction}
\label{sec1} \vspace{-0.5cm}
Nowadays the hetero-modulus ceramic composites with different mechanical properties are one from the main directions of the development of new technical materials. The most common material is the lead zirconate titanate PbZr$_{1-x}$Ti$_x$O$_3$ (PZT) family which is well-known to have high piezoelectric strain constant, electric permittivity and electromechanical coupling \cite{1}. It is too early to generalize conclusions about using PZT in technical applications. They are very wide, starting from crystals, across ceramics, fibres, ending on nanostructures and nanoparticles. PZT thin fibres have recently attracted considerable attention due to their applications in precision Microelectromechanical Systems (MEMS) or integrated electro-ceramic MEMS platforms \cite{2}. They have a large potential in sensoric, actuatoric and ultrasonic transducer applications \cite{3}. Fibres PZT have a potential for utilization in high performance hydrophones and ultrasonic transducer applications because the fine fibrous geometry offers the possibility of composite fabrication, where damping and reinforcement is combined \cite{4}. PZT fibres are used to form active elements within a functional devices as the active fibre composite~(AFC)~\cite{5}.
	
 For the better use of PZT advantages the further theoretical and experimental investigations are highly required. \textit{Ab initio} calculations offer one of the most powerful tools for carrying out theoretical studies of these properties. First \textit{ab initio} calculations for PbZr$_{0.5}$Ti$_{0.5}$O$_3$ were performed by G. S\'{a}ghi-Szab\'{o} \textit{et al}. \cite{6}. Apart from the stable ferroelectric ground state they determined the bulk spontaneous polarization, dynamical charges, and piezoelectric stress tensor elements.
Piezoelectric parameters achieve maximal values near the morphotropic boundary. Experimental results show that up to $x = 0.5$ of the Zr content these fibres maximally contain about 90\% the volume fraction of a tetragonal phase and 10\% a rhombohedral phase in a volume content of PZT \cite{7}. For this reason in this paper we have analyzed  structural, elastic, electronic properties of the tetragonal phase of  PbZr$_{0.5}$Ti$_{0.5}$O$_3$  from \textit{ab initio} calculations, near the morphotropic boundary i. e. close to  $x = 0.53$ \cite{1}. 

This paper is organized as follows: the method of calculations is described in Section 2. In Section 3 the results and discussions of the structural, elastic and electronic properties as the results of the calculations are presented. Finally, the conclusions and remarks are given in Section \ref{sec4}. 

\section{Method of calculations}
\label{sec2}
\begin{figure}[t]
\centering
\includegraphics[width=0.6\textwidth]{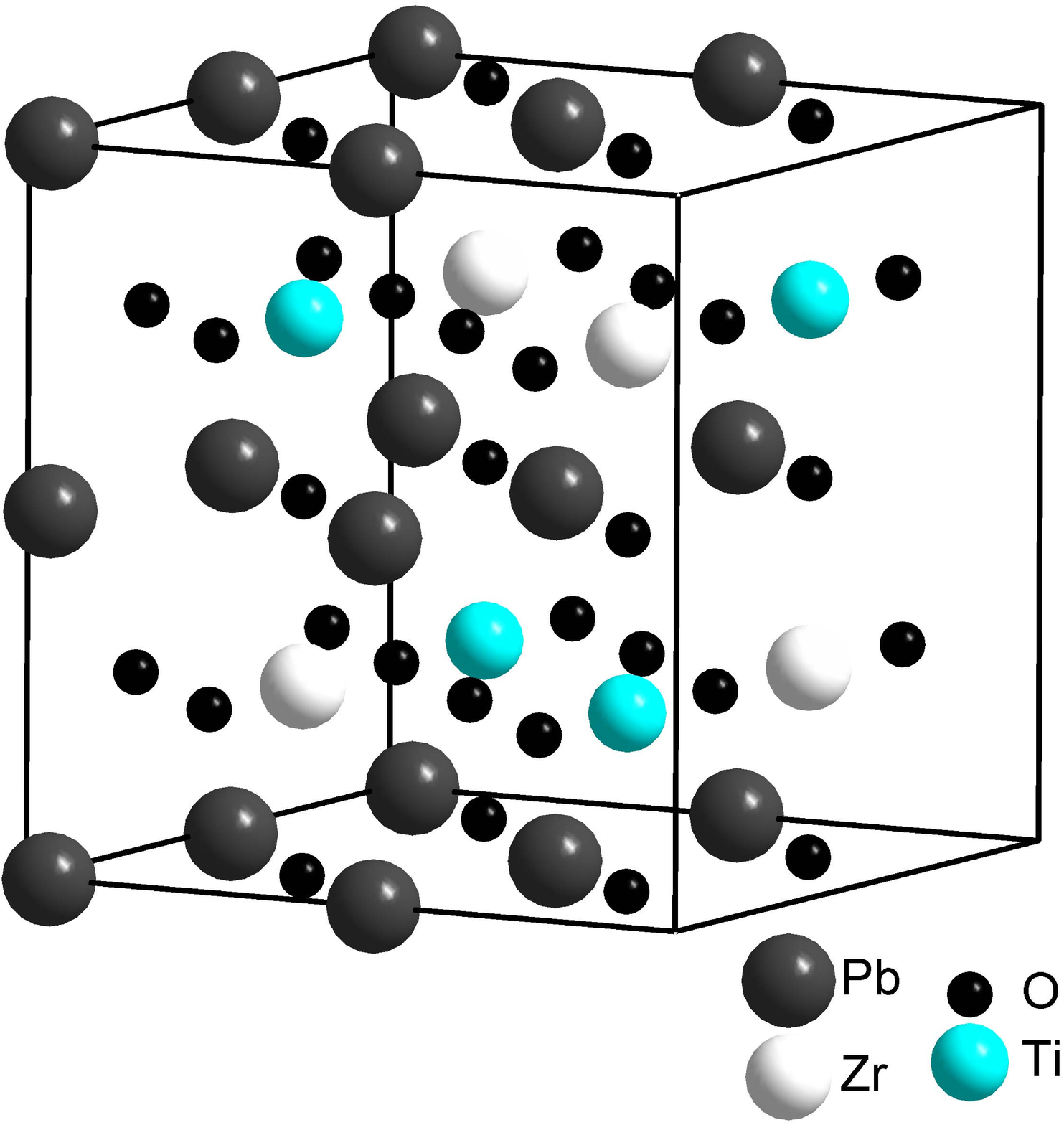}
\caption{The model unit cell with the positions of atoms in PbZr$_{0.5}$Ti$_{0.5}$O$_3$. The size of wheels in the figure is without meaning.}
\label{fig1}
\end{figure}

Density functional theory (DFT) \cite{8} calculations were performed in the generalized-gradient approximation (GGA) \cite{9}. The localized atomic-like orbitals for the basis set expansion were used as implemented in the SIESTA software package \cite{10,11}. After initially carried out tests for different possibilities the plane wave energy cutoff for the basis set at 400 Ry and a $2\times 2\times 2$ Monkhorst-Pack grid for the Brillouin zone integration were chosen \cite{12}. The unit cell containing 40 atoms \cite{13}, \cite{14} is shown in Figure \ref{fig1}. Positions of Zr and Ti atoms were chosen alternately, as should be in the ideal crystal. All positions between atoms of the each one element are equivalent.

 In order to achieve better performance, the most non-valence electrons were replaced by effective norm-conserving Troullier-Martins pseudopotentials in the Perdew-Burke-Ernzerhof (PBE) parametrization \cite{15}. In contrast, the electrons 
Ti of 
4\textit{p}$^0$
3\textit{d}$^2$ 
4\textit{f}$^0$, 
Zr 
5\textit{s}$^2$
5\textit{p}$^0$  
4\textit{d}$^2$ 
4\textit{f}$^0$, 
Pb 
6\textit{s}$^2$ 
6\textit{p}$^2$ 
6\textit{d} $^{0}$ 
5\textit{f}$^{0}$ 
and 
O  
2\textit{s}$^2$ 
2\textit{p}$^4$ 
3\textit{d}$^0$ 
4\textit{f}$^0$ 
were treated as valence electrons.

The geometric optimization of atomic positions and lattice parameters with respect to the ground state energy was performed using a conjugate-gradients method. The obtained values of lattice parameters, in the relaxation procedure, are very close to the experimental ones \cite{16}. They are presented in Table \ref{table1}.    

\begin{table}[t]
\centering
\caption{Values of calculated and experimental lattice parameters of PbZr$_{0.5}$Ti$_{0.5}$O$_3$}
\begin{tabular}{lcc}
Lattice parameters 	&	Present Work	&	Experiment \cite{14}\\
$a$ (\AA)      		&			4.120		&			4.042\\
$c$ (\AA)				&			4.122		&			4.127\\
\end{tabular}
\label{table1}
\end{table}

\section{Results of calculations}
\label{sec3}
\subsection{Elastic properties}
\label{sec31}
\begin{figure}[t]
\centering
\includegraphics[width=0.9\textwidth]{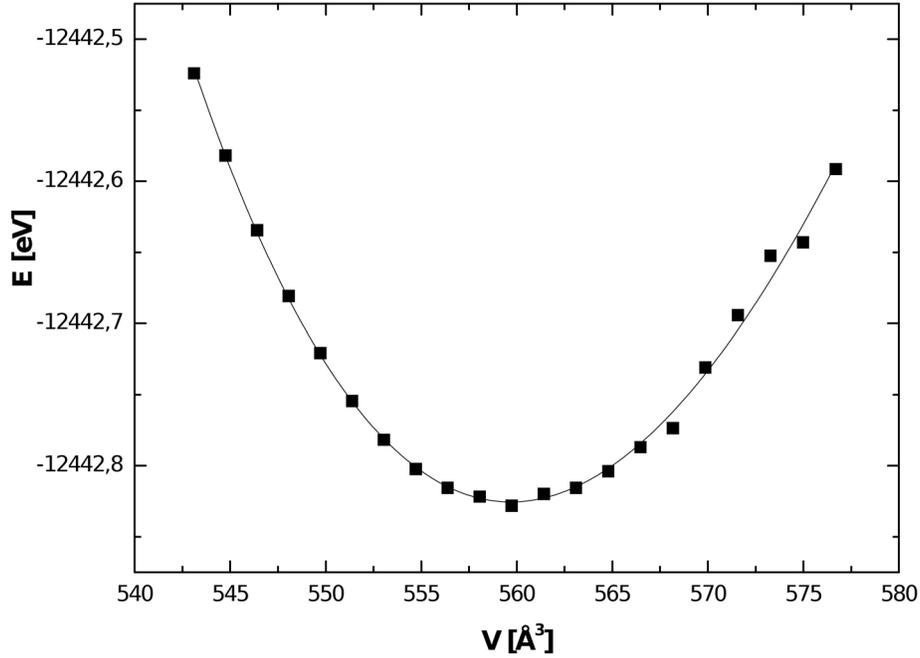}
\caption{The dependence of the total energy on the unit cell volume of PbZr$_{0.5}$Ti$_{0.5}$O$_3$}
\label{fig2}
\end{figure}

We have calculated elastic properties both  the uniform volume compressibility (the bulk modulus) as well as elastic constants, which describe different directional deformations. The bulk modulus was calculated by fitting the dependence of the total energy of the unit cell volume to a Birch-Murnaghan equation of state \cite{17} (see Fig. \ref{fig2}).  The dependence of the energy on volume was obtained by series of calculations, where the unit cell volume was varied between about  -3 \%  and +3 \%. Such relative volume change is equivalent to the hydrostatic pressure in the range from about -4 GPa to +6 GPa. From Fig. \ref{fig2} one can see that the minimum of the total energy E$_0$ = - 12443  eV (unphysical) occurs for V$_0$=560~\AA$^3$.  The calculated bulk modulus amounts to B$_0$ =170 GPa. 

\begin{figure}[t]
\centering
\includegraphics[width=1.0\textwidth]{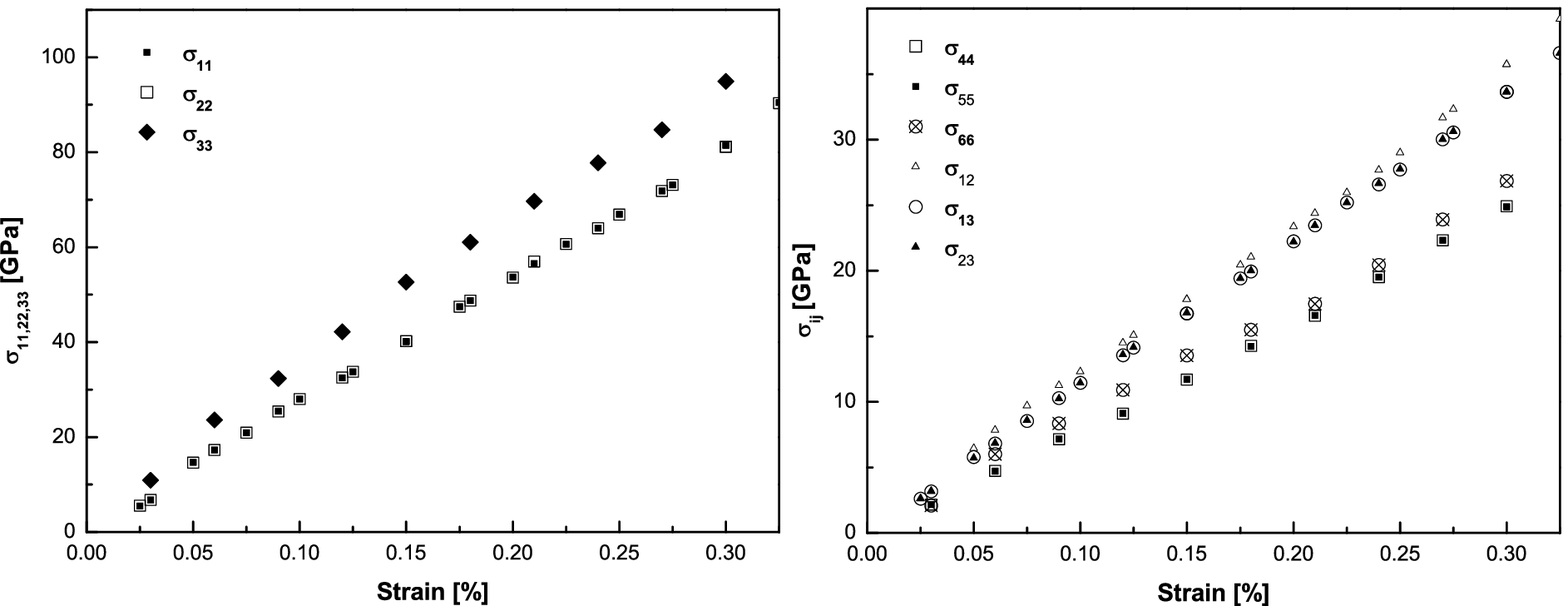}
\caption{Calculated directional stress-strain dependences of $\sigma_{11}$, $\sigma_{22}$, $\sigma_{33}$ (a) and (b) 
$\sigma_{44}$, $\sigma_{55}$, $\sigma_{66}$, $\sigma_{12}$ , $\sigma_{23}$, and $\sigma_{13}$ for  PbZr$_{0.5}$Ti$_{0.5}$O$_3$}
\label{fig3}
\end{figure}

The dependence of $\sigma_{11}, \sigma_{22}, \sigma_{33}, \sigma_{44},  \sigma_{55}, \sigma_{66}, \sigma_{12}, \sigma_{23},$ and $\sigma_{13}$ on the strain was calculated in the range 0.03\%-0.3\% by applying small strains (for higher deformations the material is destroyed) in selected directions with values of the corresponding stress tensor components from the Kohn-Sham total energy [8]. The stresses $\sigma_{11}$, $\sigma_{22}$ and $\sigma_{44}$, $\sigma_{55}$ are equal in pairs due to  the tetragonal symmetry in PbZr$_{0.5}$Ti$_{0.5}$O$_3$. Results are presented in Figs \ref{fig3}(a) and \ref{fig3}(b). 

Values of $\sigma_{11}$  and  $\sigma_{22}$ are about 15\% lower than $\sigma_{33}$. It means that in the $c$ direction the tetragonal PbZr$_{0.5}$Ti$_{0.5}$O$_3$ is harder than in the ($a,b$) plane. The stress $\sigma_{12}$ and $\sigma_{13}$ describe properties of the PbZr$_{0.5}$Ti$_{0.5}$O$_3$ in the ($a,b$) and ($a,c$) plane, respectively. There are large differences between them and other elastic constants. The stress  $\sigma_{44}$ corresponds to the shear deformation in the ($b,c$) plane. All elastic constants for the shear deformation are significantly smaller than the elastic constants for  the uniaxial deformation. During calculations we have observed almost the same behaviour for the compression and the elongation in the dependence of these elastic constants on the applied strain. Similarly to an experiment \cite{18} one can observe the nonlinear behaviours for small strains near 0.05\% and next near 0.20\%. The value 0.25\% of the strain is critical  - about this value  real crystals are destroyed. This fact is in very good agreement with the experimental one \cite{18}.
 \begin{figure}[t]
\centering
\includegraphics[width=0.7\textwidth]{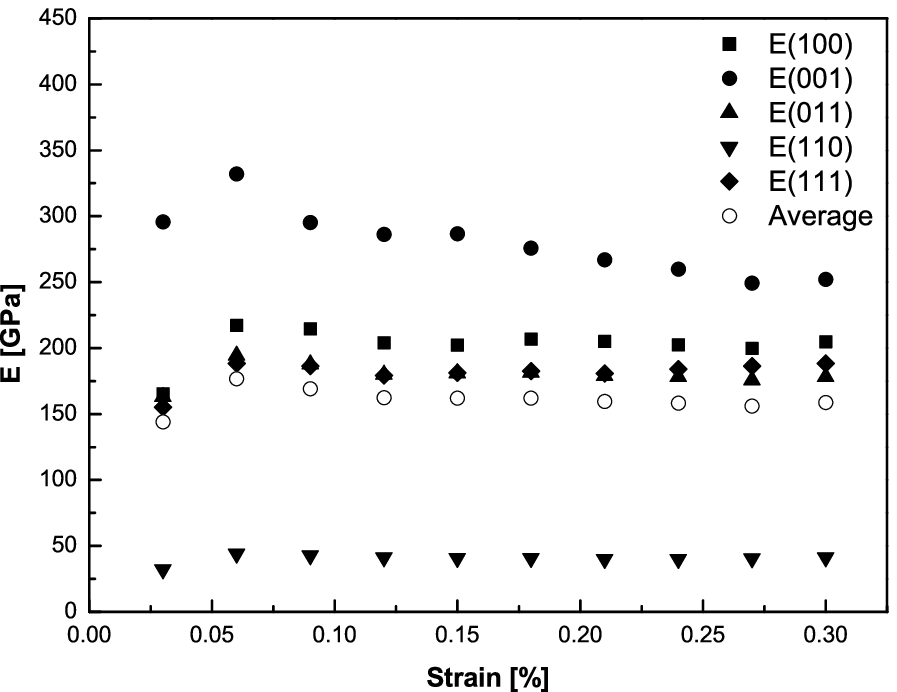}
\caption{Calculated values for the average Young's modulus as the  function of the strain in PbZr$_{0.5}$Ti$_{0.5}$O$_3$}
\label{fig4}
\end{figure}
Basing on these results for the strain range of 0.03\%-0.3\%  we have calculated the respective elastic constants, the directional Young's modulus E and average values of Young's modulus \cite{19}. In Fig.\ref{fig4} the dependence of the average Young's modulus  E on the strain for PbZr$_{0.5}$Ti$_{0.5}$O$_3$ is shown. Between strain  values of  0.07\% and 0.15\%  there exists only very  short region, where the Hooke law is valid. The highest values of the Young modulus for the \textit{c}-direction are found, and the smallest ones arose for the (\textit{a} , \textit{b})-plane.

\subsection{Electronic structure}\label{sec32}
\begin{figure}[h!]
\centering
\includegraphics[width=0.6\textwidth]{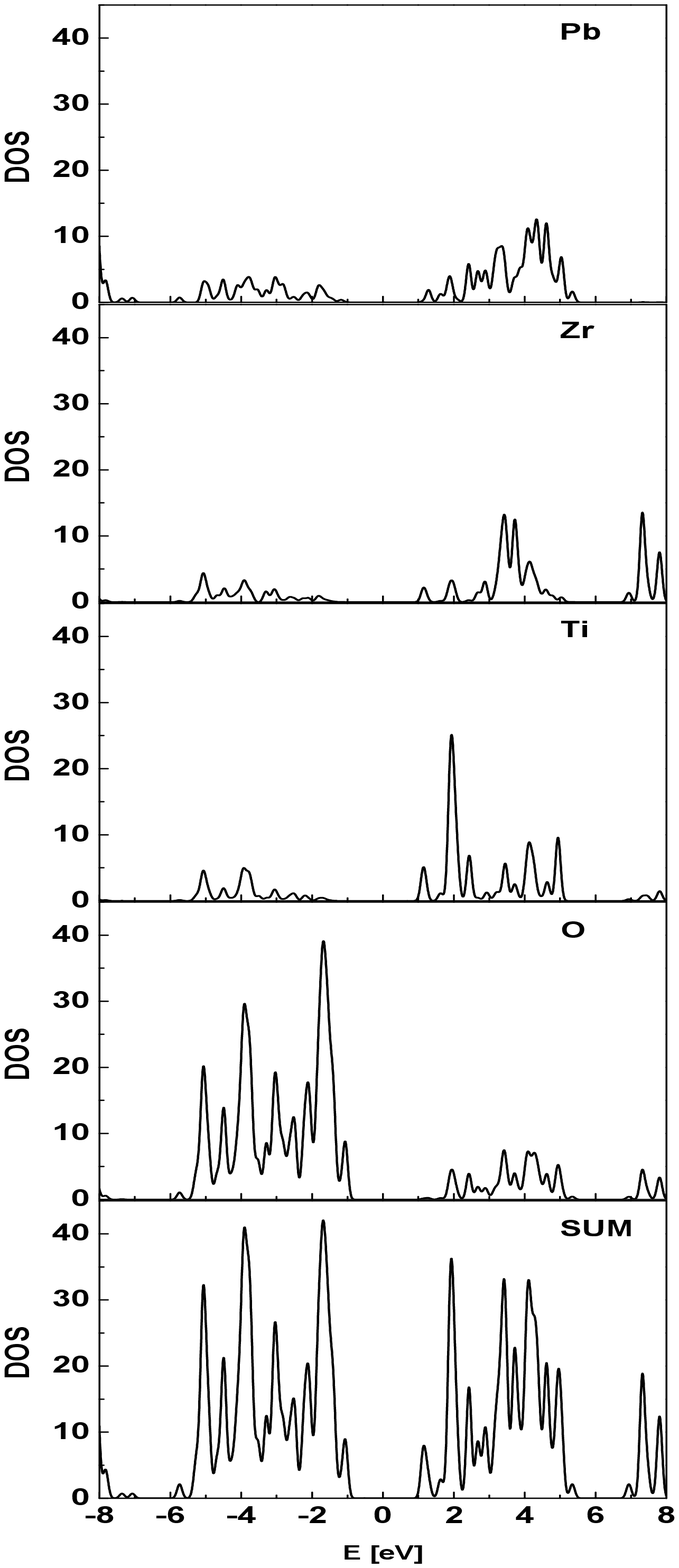}
\vspace{-0.5cm}
\caption{Total density of states for  the tetragonal phase PbZr$_{0.5}$Ti$_{0.5}$O$_3$}
\label{fig5}
\end{figure}

The calculated total density of states for  the tetragonal phase PbZr$_{0.5}$Ti$_{0.5}$O$_3$  is presented in Figure \ref{fig5}.
 The value of a band gap amounts to E$_g$ = 2.1 eV. It must be noticed, that a width of the gap depends on a manner of the reading from the diagram - how many states we mean as empty or occupied by electrons. Probably states with very small number of states are not registered in real conditions. If we consider the level of an occupancy of the middle states as approximately equal to zero we will obtain the energy gap 3.5 eV - this result agree with the experimental one very well \cite{20}. Moreover calculations are made for ideal crystal at 0 K but measurements were performed at room temperature. In the SIESTA program there are not possible to include the different temperature into account. 
The net atomic charges, collected in Table \ref{table2}, indicate the substantial ionicity of the atoms. The net charges were calculated by the integration of the charge density over a sphere of the selected radius. They are the real charges in the nodes of  the lattice. The integration radii were taken as 1.25 a.u., 1.5 a.u., 1.7 a.u., and 2.5 a.u. for Ti,  Zr, Pb and  O atoms, respectively. These radia are about a half of the bond length between respective atoms. There  exist differences between the calculated and nominal charges of atoms. The biggest ones for Pb ions being about 50\% larger than +2 are supposed to be due to a substantial hybridization between the O orbitals with valence orbitals of  other ions. The most likely they are caused by hybridization between O 2\textit{p} and Pb 6\textit{s} orbitals  \cite{20},  \cite{21}. This fact has its grounds in a partial density of states PDOS showed in Fig. \ref{fig5}. For such effect is responsible the splitting between the frequencies of the  longitudinal optical (LO) and transverse optical (TO) phonons  \cite{xxx}. Existence of Pb$^{+3}$ ions in PZT is experimentally confirmed by EPR measurements \cite{20}. This mechanism is responsible for a great sensitivity of the ferroelectricity to the domain structure and boundary conditions, especially for very good ferroelectric properties of PZT.

\begin{table}[t]
\centering
\caption{Calculated net ionic charges in PbZr$_{0.5}$Ti$_{0.5}$O$_3$}
\begin{tabular}{lcccc}
										&	Pb$^{II,IV}$						&	Zr$^{IV}$	&	Ti$^{III,IV}$		&	O$^{-II}$		\\
Radius of integration (a.u.)	&	1.7									&	1.5			&	1.25					&	2.5				\\
Calculated atomic charge 		&	3.25									&	3.53			&	3.43					&	-1.24				\\
\end{tabular}
\label{table2}
\end{table}

\section{Discussion and conclusions}
\label{sec4}
From the analysis of the stress-strain directional dependences and the Young modulus-strain dependency one can see that in PbZr$_{0.5}$Ti$_{0.5}$O$_3$ near morphotropic boundary the Hooke law is carried out only on the range of the strain 0.07\%-0.15\%. Values of directional Young's modulus are approximately constant in this range. It seem to be  typical for the stress-strain behaviour of PbZr$_{0.5}$Ti$_{0.5}$O$_3$  fibres as has been written in Refs \cite{16, 18, 22}. Theoretical results are over estimate in comparison with the  experimental ones - it is supposed to be due to defects occurring even in the best real crystals (calculations have been performed for the stoichiometric ideal crystal structure and the temperature at 0 K). The experimental value of  the Young's modulus along the radius for ceramic PZT fibres are about 108 GPa-120 GPa in the room temperature \cite{23}. The real value always is lower than this theoretical one at 0 K because of presence some defects and higher temperatures.   
 
The same effect has been observed in \textit{ab initio} calculations for other perovskite crystal \cite{24,25}. Knowing from theoretical calculations that point defects cause decreasing almost by a half of calculated elastic constants of the ideal crystal it is necessary to compare the calculated values of these parameters for the defected crystals with the experimental values of elastic constants for tetragonal PZT measured at low temperatures.
%\vspace{-0.5cm}
\section*{Acknowledgements} 
%\vspace{-0.5cm}
I would like to thank David M. Nalecz and Kacper Druzbicki for fruitful discussions. I acknowledge the CPU time allocation at Academic Computer Centre CYFRONET AGH in Cracow. This work was supported by the grants of MNiSW/SGI3700/AP/025/2007 and the PL-Grid Infrastructure.



%% References


\bibliographystyle{model1a-num-names}
\bibliography{<your-bib-database>}

\begin{thebibliography}{00}


\bibitem{1} A. G. Ye (ed.), Handbook of dielectric, piezoelectric and ferroelectric materials, Woodhead Publishing Limited, Cambridge, 2008.
\bibitem{2} N. Setter, Electroceramics-based MEMS Fabrication-Technology and Applications, Springer, New York, 2005.
\bibitem{3} K. Uchino, Ferroelectric Devices, Marcel Dekker Inc., New York, 2000.
\bibitem{4} S. Yoshikawa, S. Ulgargaraj, P. Moses, J. Witham, R. Meyer and T. Shrout, Pb(Zr,Ti)O$_3$ [PZT] fibres-fabrication and measurement methods, J. Int. Mat. Syst. Struct. 6  (1995), pp. 152$-$158.                                                                                                                                                                                        \bibitem{5} A. A. Bent and N. W. Hagood, Piezoelectric fibre composites with interdigitated electrodes, Int. Mater. Syst. Struct. 8 (1997), pp.  903$-$919.
\bibitem{6} G. S\'{a}ghi-Szab\'{o}, R. E. Cohen and H. Krakauer, First-principles study of piezoelectricity in tetragonal PbTiO$_3$ and  Pb(Zr$_{0.5}$Ti$_{0.5}$)O$_3$ , Phys. Rev. B 59 (1999), pp. 12771$-$12776.
\bibitem{7} R. Steinhausen, T. Hauke, H. Beige, W. Watzka, U. Lange, D. Sporn, S. Gebhardt and A. Sch\"{o}necker,  Properties of fine scale piezoelectric PZT fibers with different Zr content, J. Eur. Ceram. Soc. 21 (2001), pp. 1459$-$1462.
\bibitem{8} W. Kohn and L. J. Sham,  Self-consistent equations including exchange and correlation effects, Phys. Rev. 140 (1965), pp. 1133$-$1138.
\bibitem{9} J. P. Perdew, J. A. Chevary, S. H. Vosko, K. A. Jackson, M. R. Pederson, D. J. Singh and C. Fiolhais,  Atoms, molecules, solids, and surfaces: Applications of the generalized gradient  approximation for exchange and correlation, Phys. Rev. B. 46 (1992), pp. 6671$-$6687.
\bibitem{10} J. Soler, E. Artacho, J. D. Gale, A. Garc\'{i}a, J. Junquera, P. Ordej\'{o}n and D. S\'{a}nchez-Portal, The SIESTA method for ab initio order-N materials simulation,  J. Phys.: Condens. Matter 14 (2002), pp. 2745$-$2779.
\bibitem{11} P. Ordej\'{o}n, E. Artacho, J. Soler,    Self-consistent order-N density-functional calculations for very large systems, Phys. Rev. B 53 (1996), pp. R10441$-$R10444.
\bibitem{12} H. J. Monkhorst and J. D. Pack, Special points for Brillouin-zone integration,  Phys. Rev. B.  3 (1976), pp. 5188$-$5192.
\bibitem{13} K. M. Rabe, C. H. Ahn, J-M. Triscone (eds.), Physics of Ferroelectrics, Springer-Verlag, Berlin Heidelberg, 2007, p.12.
\bibitem{14} A. Boonchun, M. F. Smith, B. Cherdhirunkorn, and S. Limpijumnong, First principles study of Mn impurities in PbTiO$_3$ and PbZrO$_3$, J. Appl. Phys. 101 (2007) pp. 043521-1$-$043521-7.
\bibitem{15} J. P. Perdew, K. Burke, and M. Ernzerhof, Generalized gradient approximation made simple, Phys. Rev. Lett. 77 (1996), pp. 3865$-$3868.
\bibitem{16} G.  A. Rossetti Jr., G. Popov, E. Zlotnikov and N. Yao, Domain structures and nonlinear mechanical deformation of soft Pb(Zr$_{x}$Ti$_{1-x}$)O$_3$ (PZT) piezoelectric ceramic fibers, Mat. Sci. and Eng. A. 433 (2006), pp. 124$-$132.
\bibitem{17} F. Birch, Finite elastic strain of cubic crystals, Phys. Rev. 71 (1947), pp. 809$-$824.                                                                                                         
\bibitem{18} R. Dittmer, F. Clemens, A. Schoenecker, U. Scheithauer, M. Rojas-Ismael, and T. Graule, Microstructural analysis and mechanical properties of Pb(Zr,Ti)O$_3$ fibers derived by different processing routes, J. Am. Ceram. Soc.  93  (2010), pp. 2403$-$2410.
\bibitem{19} J. F. Nye, Physical Properties of Crystals: Their Representation by Tensors and Matrices, Oxford University Press, USA, 1985. 
\bibitem{20} J. Robertson, W. L. Warren, and B. A. Tuttle, Band states and shallow hole traps in Pb(Zr,Ti)O$_3$ ferroelectrics, J. Appl. Phys. 77 (1995) pp. 3975$-$3980.
\bibitem{21} W. L. Warren, J. Robertson, D. Dimos, B. A. Tuttle, G. E. Pike, and D. A. Payne, Pb displacements in Pb(Zr,Ti)O$_3$ perovskites, Phys. Rev. B 53 (1995) pp. 3080$-$3087.
\bibitem{xxx} W. Zhong, R.D. King-Smith, and D. Vanderbilt, Giant LO-TO Splittings in Perovskite Ferroelectrics, Phys. Rev. Lett. 72 (1994) pp. 3618-3621
\bibitem{22} X. Kornmann and C. Huber, Microstructure and mechanical properties of PZT fibres, J. Eur. Ceram. Soc. 24 (2004), pp. 1987$-$1991.                                                                                   
\bibitem{23} L. Kozielski, M. Piechowiak, D. Czekaj, A. Lisinska-Czekaj, A. Smalarz, F. Clemens, and R. Nowak, Mechanical examination of PZT microfibre defects structure, Phase Trans. 81 (2008), pp.1081$-$1088.
\bibitem{24} R. Bujakiewicz-Koronska and Y. Natanzon, First principles calculations of elastic constants for defected Na$_{1\slash 2}$Bi$_{1\slash 2}$TiO$_3$ , Integr. Ferroelectrics. 108 (2009), pp. 21$-$36.
\bibitem{25} R. Bujakiewicz-Koronska and Y. Natanzon,  Determination of elastic constants of Na$_{0.5}$Bi$_{0.5}$TiO$_3$ from ab initio calculations, Phase Trans. 81 (2008), pp.1117$-$1124.
\end{thebibliography}



\end{document}